\documentclass[numberedappendix,appendixfloats,iop,apj]{emulateapj}

\usepackage{rotating}
\newcommand{\mbh}{$M_{\rm BH}$}

\newcommand{\ms}{$M_{\rm sph}$}

\newcommand{\ls}{$L_{\rm sph}$}

\newcommand{\s}{$\sigma$}
\newcommand{\msun}{$M_{\odot}$}

\usepackage[usenames,dvipsnames]{color}
\usepackage{times, apjfonts}
\usepackage{amsmath, amsthm, amssymb}
\usepackage{graphicx, epsfig}
\usepackage{natbib}
\usepackage{epstopdf} 
\usepackage{CJK} 
\usepackage[colorlinks=true, anchorcolor=blue, linkcolor=blue, urlcolor=Fuchsia, citecolor=NavyBlue, breaklinks=true]{hyperref}
\usepackage[hyphenbreaks]{breakurl}

\begin{document}

\title{A LOCAL BASELINE OF THE BLACK HOLE MASS SCALING RELATIONS FOR
  ACTIVE GALAXIES. III.\\
THE \mbh-\s~RELATION}

\shorttitle{A LOCAL BASELINE OF THE \mbh~SCALING RELATIONS FOR ACTIVE GALAXIES. III.}
\shortauthors{Bennert et al.}

\author{Vardha N. Bennert\altaffilmark{1},
Tommaso Treu\altaffilmark{2}, 
Matthew W. Auger\altaffilmark{3}, 
Maren Cosens\altaffilmark{1},
Daeseong Park\altaffilmark{4},
Rebecca Rosen\altaffilmark{1,5},
Chelsea E. Harris\altaffilmark{6},
Matthew A. Malkan\altaffilmark{7},
Jong-Hak Woo\altaffilmark{8}}

\altaffiltext{1}{Physics Department, California Polytechnic State
  University, San Luis Obispo, CA 93407, USA; vbennert@calpoly.edu; mcosens@calpoly.edu}

\altaffiltext{2}{Department of Physics, University of California, Santa
Barbara, CA 93106, USA; tt@physics.ucsb.edu; Sloan Fellow, Packard Fellow;
present address: Department of Physics and Astronomy, University of California,
Los Angeles, CA 90095, USA}

\altaffiltext{3}{Institute of Astronomy, Madingley Road, Cambridge CB3 0HA, UK; mauger@ast.cam.ac.uk}

\altaffiltext{4}{National Astronomical Observatories, Chinese Academy of Sciences, Beijing 100012, China; daeseong.park@nao.cas.cn}

\altaffiltext{5}{Present address: AstroCamp, Idyllwild, CA 92549, USA; rosenrr@prodigy.net}

\altaffiltext{6}{Lawrence Berkeley National Laboratory, Berkeley,
  California 94720, USA; ChelseaHarris@lbl.gov}

\altaffiltext{7}{Department of Physics and Astronomy, University of California,
Los Angeles, CA 90095, USA; malkan@astro.ucla.edu}

\altaffiltext{8}{Department of Astronomy, Seoul National University, Korea; woo@astro.snu.ac.kr}

\shortauthors{Bennert et al.}

\begin{abstract}
We create a baseline of the black hole (BH) mass (\mbh) --
stellar-velocity dispersion ($\sigma$) relation for active galaxies,
using a sample of 66 local ($0.02<z<0.09$) Seyfert-1 galaxies,
selected from the Sloan Digital Sky Survey (SDSS). Analysis of
SDSS images yields AGN luminosities free of host-galaxy contamination,
and morphological classification. 51/66 galaxies have spiral
morphology. 28 bulges have S{\'e}rsic index $n<2$ and are considered
candidate pseudo bulges, with eight being definite pseudo bulges based on
multiple classification criteria met. Only 4/66
galaxies show signs of interaction/merging. High signal-to-noise ratio
Keck spectra provide the width of the broad H$\beta$ emission line
free of FeII emission and stellar absorption. AGN luminosity and
H$\beta$ line widths are used to estimate \mbh. The Keck-based
spatially-resolved kinematics is used to determine stellar-velocity
dispersion within the spheroid effective radius ($\sigma_{\rm spat,
reff}$).  We find that \s\, can vary
on average by up to 40\% across definitions commonly used in
the literature, emphasizing the importance of using self-consistent
definitions in comparisons and evolutionary studies. The \mbh-$\sigma$
relation for our Seyfert-1 galaxy sample has the same intercept and
scatter as that of reverberation-mapped AGNs as well as that of
quiescent galaxies, consistent with the hypothesis that our single
epoch \mbh\ estimator and sample selection do not introduce
significant biases. Barred galaxies, merging galaxies, and those
hosting pseudo bulges do not represent outliers in the
\mbh-$\sigma$ relation. This is in contrast with previous work, 
although no firm conclusion can be drawn on this matter due to the
small sample size and limited resolution of the SDSS images.
\end{abstract}


\section{INTRODUCTION}
\label{sec:intro}
The discovery of relations between the mass of the central
supermassive black hole (BH) and its host galaxy properties
such as spheroid luminosity \ls~\citep{kor95},
spheroid mass \ms~\citep[e.g.,][]{mag98},
and spheroid stellar velocity dispersion
\s~\citep[e.g.,][]{geb00,fer00}
has sparked a flood of observational studies pertaining both to the local
Universe \citep[e.g.,][]{mer01,tre02,mar03,har04,fer05,gre06,gra07,gue09,ben11a,korb11,kor11,san11,mcc11,gra11,bei12,gra13}
and cosmic history
\citep[e.g.,][]{tre04,tre07,pen06a,pen06b,woo06,woo08,sal07,rie09,jah09,ben10,dec10,mer10,ben11b};
for a recent review see \citet{kor13}.
In particular the evolution with redshift of these correlations
constrains theoretical interpretations and provides important
insights into their origin \citep[e.g.,][]{cro06,rob06,hop07},
by probing whether BHs and their host galaxies
are constantly on tight correlations 
through a feedback mechanism that controls
their mutual growth \citep[e.g.,][]{kau00,vol03,cio07,hop08},
or whether the local relations are an end product 
of the more dramatic and stochastic process
of galaxy merging in the hierarchical assembly of \mbh~and stellar
mass \citep[e.g.,][]{pen07,jah11}.

Measuring \mbh~based on 
spatially resolving the BH's gravitational sphere of influence via stellar kinematics
\citep[e.g.,][]{van98,geb00}, gaseous kinematics \citep[e.g.,][]{fer96,mar01}
or maser emission \citep[e.g.,][]{her05,kuo11} is restricted to
galaxies in the local Universe.  The only way to probe the evolution
of the scaling relations is to rely on active galaxies, thought to
represent an integral phase in the evolution of galaxies during which
the BH is growing through accretion, resulting in the luminous galaxy
center known as Active Galactic Nucleus (AGN).  For AGNs at
cosmological distances, \mbh~is estimated through application of
the ``virial method'' \citep[e.g.,][]{wan99}.  In this
method, it is assumed that gas clouds in the broad-line region (BLR)
orbiting the BH in close proximity follow the gravitational field of
the BH.  In a time-consuming process called reverberation mapping
\citep[e.g.,][]{wan99, kas00, kas05, ben06,ben13},
the delayed response of broad emission lines to changes in the AGN continuum
is translated into a size of the BLR. 
When combined with the width of the broad emission lines (i.e.,~velocity of the gas clouds),
the mass of the BH can be estimated, making use of the ``virial coefficient''
that depends on the gas kinematics and geometry.
Assuming that broad-line AGNs and quiescent galaxies follow
the same \mbh-\s~relation
\citep[as probed in several studies, e.g.,][and references therein]{woo13},
this coefficient has been traditionally obtained by matching their \mbh-\s~relations
\citep[e.g.,][]{onk04, gre06, woo10,par12,woo13,woo15}.
More recently, by modeling reverberation-mapped data directly and
constraining geometry and kinematics of the BLR, \mbh~has been
estimated for individual objects independent of a virial coefficient
\citep{bre11,pan11,pan12,pan14}.  A secondary method called the
``single-epoch method'' makes use of an empirical correlation found from
reverberation mapping that directly relates the BLR size to the AGN
continuum luminosity, to allow the estimation of \mbh~from a single
spectrum \citep{ves02,woo02,ves06,mcg08}. 

While the majority of evolutionary studies point toward
a scenario in which BH growth precedes spheroid assembly
\citep[e.g.,][]{wal04,tre04,woo06,shi06,mcl06,pen06a,pen06b,tre07,sal07,wei07,rie08,woo08,rie09,gu09,jah09,dec10,mer10,ben10,ben11b},
no consensus has been reached on the interpretation of the evolutionary
studies \citep[see e.g.,][especially concerning the role of scatter, observational bias, and selection effects]{sch13,vol11,sch14} and, ultimately, the origin of the \mbh~scaling relations.
A key toward understanding the \mbh~scaling relations
may lie in understanding the local relations for active galaxies
and systematic effects in the analysis.
First, all conclusions about evolution
of these relations hinge on understanding the slope and scatter of
local relations, especially those involving broad-line AGNs --- the class
of objects targeted by high-redshift studies, by necessity.  
Second, while reverberation-mapped AGNs benefit from
smaller \mbh~errors, their selection based on sufficient AGN
variability may introduce biases \citep[e.g.,][and references therein]{woo13}.
Third, investigating the dependence of scaling relations on additional
parameters, such as the amount of nuclear
activity and the detailed properties of the host galaxies, is vital
to understanding the physical origins of galaxies.
For example, while the spheroid has been traditionally identified
as the fundamental driver for \mbh, 
there have been studies that point towards tighter correlations of \mbh~with the
total host-galaxy light or stellar mass
\citep{ben10,ben11b,jah09,lae14}.
A related open question is 
the role of pseudo bulges.
Late-type galaxies are often known to host pseudo bulges,
characterized by nearly exponential light profiles, ongoing star
formation or starbursts, and nuclear bars.  It is generally
believed that they have evolved secularly through dissipative
processes rather than mergers \citep{cou96,kor04}.
Conversely, classical bulges are thought
of as centrally concentrated, mostly red and quiescent,
merger-induced systems.
It is unclear how BHs would grow within pseudo bulges and how
their masses could be related: some authors find that pseudo bulges
correlate with \mbh~\citep[e.g.,][]{kor01,gu09},
while others propose either the opposite
\citep{hu08,gre10,korb11}
or at least that both the \mbh-\s~and \mbh-\ls~relations are not
obeyed simultaneously
\citep{gre08,now10}.

We here present the results of a program 
aimed at addressing these questions by building upon a robust and unique baseline of $\sim$100
local (0.02 $\le$ $z$ $\le$ 0.09) Seyfert-1 galaxies selected from the 
Sloan Digital Sky Survey (SDSS)
(\mbh$>10^{7}$M$_{\odot}$) for the study of the \mbh~scaling relations. 
The homogeneous selection of our sample based on emission lines
is disjoint from the reverberation-mapped AGNs and allows us to probe selection effects
in the reverberation-mapped AGN sample which serves as a \mbh~calibrator for the entire Universe.
Moreover, our selection is similar to high-redshift samples.
Combining high-quality long-slit Keck/LRIS spectra  
with archival multi-filter SDSS images yields four different
fundamental scaling relations.
Results for a pilot sample of 25 objects have been presented by
the first paper in the series
\citep[][hereafter Paper I]{ben11a}. 
Spatially-resolved \s~measurements for the full sample have been published by
\citet{har12}[hereafter Paper II].

In this paper, we focus on the \mbh-\s~relation.
We have obtained high S/N spatially-resolved
long-slit spectra and measured both aperture \s~as well as spatially
resolved \s~(Paper II).
Here, we derive spatially resolved \s~within the bulge
effective radius and compare it to different definitions of stellar velocity
dispersion used in the literature.

The paper is organized as follows.
We summarize sample selection, observations, 
and data reduction in Section~\ref{sec:sample}.
Section~\ref{sec:derived} describes the derived quantities,
such as host-galaxy properties derived from surface photometry, stellar velocity dispersion, 
and \mbh.
In Section~\ref{sec:final}, we describe our final sample as well as comparison samples drawn from literature,
consisting of local quiescent galaxies and reverberation-mapped AGNs.
We present and discuss our results in Section~\ref{sec:results}.
We conclude with a summary in Section~\ref{sec:summary}.
In Appendix~\ref{appendix1}, we show fits to the broad H$\beta$ emission line for 
a total of 79 objects for which we measured \mbh.
Appendix~\ref{appendix2} summarizes notes for a few individual objects.

Throughout the paper, we assume
a Hubble constant of $H_0$ = 70\,km\,s$^{-1}$\,Mpc$^{-1}$,
$\Omega_{\Lambda}$ = 0.7 and $\Omega_{\rm M}$ = 0.3. 

\section{SAMPLE SELECTION, OBSERVATIONS, AND 
DATA REDUCTION}
\label{sec:sample}
Sample selection, observations, and data reduction
were described in detail in Paper I \& II of the series,
and are only summarized here briefly, for convenience.
Details for the full sample of 103 objects
are listed in Paper II (Tables 1\&2).

The sample was selected from the SDSS Data
Release (DR) 6 following these criteria:
(i) \mbh$>$10$^7$\msun~as estimated based on optical luminosity
and Full-Width-at-Half-Maximum (FWHM) of the broad H$\beta$ line;
(ii) redshift range 0.02 $<$ $z$ $<$ 0.09 to measure stellar kinematics
via the CaII triplet line in the optical and to ensure that the
objects are well resolved. 
A total of 103 objects were observed
between 2009 January and 2010 March with 
the Low Resolution Imaging Spectrometer (LRIS) 
at Keck I using a 1$\arcsec$ wide long slit, aligned with the
host galaxy major axis as determined from SDSS (``expPhi\_r''). 
The D560 dichroic (for data taken in 2009) 
or the D680 dichroic (for data taken in 2010) was used, 
the 600/4000 grism in the blue, and the 831/8200 grating in the red
with central wavelength 8950\AA,
resulting in an instrumental resolution of
$\sim$90\,km\,s$^{-1}$ in the blue and $\sim$45\,km\,s$^{-1}$ in the
red.
A table with sample and observation details can be found in Paper II (Table 1).

The data were reduced following standard reduction steps 
including bias subtraction, flat fielding, and cosmic-ray rejection. 
Wavelengths were calibrated using arc lamps in the blue spectrum
and sky emission lines in the red spectrum.
Telluric absorption correction and relative flux calibration was performed
using AOV Hipparcos stars.

All objects were covered by the VLA FIRST (Faint Images of the Radio Sky at Twenty-cm) 
survey\footnote{See VizieR Online Data Catalog, 8071 \citep{bec03}}, but only 32 have counterparts within a radius of 5\arcsec.
Out of these, 21 are listed in \citet{raf09} with only two being radio-loud. Thus, the majority of our objects are radio-quiet.

\section{DERIVED QUANTITIES}
\label{sec:derived}
To derive surface photometry, stellar-velocity dispersion, and \mbh,
we followed the same procedures as outlined
in Paper I \& II.
We here briefly summarize the procedure and results.

\subsection{Surface Photometry}
\label{subsec:photometry}
In Paper I, we described in detail an image analysis code
``Surface Photometry and Structural Modeling of Imaging Data''
(SPASMOID) designed to
allow for simultaneous fitting of multi-filter images with arbitrary constraints
between the parameters in each band \citep{ben11a,ben11b}.
This joint multi-wavelength analysis enables a much more powerful
disentanglement of the nuclear and host-galaxy components than
using single-band imaging alone (e.g., using GALFIT, \citealt{pen02}).
The bluer bands provide a robust measurement of the normalisation of the
nuclear flux while the redder data exploit the more favorable contrast
between the AGN and the host galaxy to constrain the morphological structure
of the latter.
The approach of simultaneously using structural and
photometric information is most successful for imaging of
AGN hosts given the presence of a bright AGN point source.
SPASMOID's reliance on a 
Markov Chain Monte-Carlo (MCMC) technique also provides realistic 
uncertainties and the ability
to explore covariances between various model parameters.

We use SPASMOID to perform surface-brightness photometry on the SDSS images,
simultaneously fitting the AGN by a point-spread function (PSF) and
the host galaxy by a combination of spheroid (S{\'e}rsic with free
index $n$, in a range between 0.7 and 4.7), and if present, disk (exponential profile), and bar
(S{\'e}rsic with index $n=0.5$, i.e., a Gaussian).
The results given in Tables~\ref{table:meas1}--\ref{table:meas2}
correspond to the  maximum a posteriori (MAP) values.
Note that this approach differs slightly from Paper I,
in which we fitted the spheroid with a single \citet{dev48} profile.
We thus here included all 25 objects from Paper I again and
ran SPASMOID on the full sample of 103 objects.
From the final sample, 11 objects were omitted
due to either image defects, bright nearby stars complicating the fit,
or no reliable fit achieved.

We use the S{\'e}rsic index to distinguish between classical bulges 
and pseudo bulges (see Section~\ref{subsec:morph}).
The spheroid radius is used to determine
the stellar velocity dispersion within that effective radius (see~\ref{subsec:dispersion}).
The PSF g'-band magnitude is corrected for Galactic
extinction (subtracting the SDSS DR7 "extinction\_g"' column), and then
extrapolated to 5100\AA, assuming a power law of the form $f_{\nu}$
$\propto$ $\nu^{\alpha}$ with $\alpha$= -0.5. 
The resulting AGN luminosity
free of host-galaxy contribution 
(except potential dust attenuation)
is used for \mbh~measurements
 (see~\ref{subsec:mbh}).
In the subsequent papers of the series (V. N. Bennert et al. 2015, in prep.), we will
discuss luminosity and stellar masses of the different components
when deriving the remaining \mbh-scaling relations.

\subsection{Stellar-velocity Dispersion and Spatially-resolved Kinematics}
\label{subsec:dispersion}
From the full sample of 103 objects, the spectra of 21 objects
did not yield a robust measurement of the stellar kinematics, 
due to dominating AGN flux and high redshift (12 objects)
or problems with the instrument (9 objects), so our final kinematic sample
consists of 82 objects (see Paper II, Table 2).
While the exclusion of 10\% of objects with faint galaxies compared to the AGN can in principle introduce a systematic effect 
in our sample, we consider this effect negligible, given the overall sample size.

For the majority of objects, broad nuclear FeII emission
($\sim$5150--5350\AA) is present and interferes with the
measurements of both $\sigma$ in the MgIb range and broad H$\beta$ width.
Thus, for those objects, a set of IZw1 templates (varying width and
strength) and a featureless AGN continuum were fitted simultaneously
and subtracted.

Stellar-velocity dispersion \s~was measured from three different
spectra regions: around CaH\&K$\lambda\lambda$3969,3934 (hereafter
CaHK), around the Mg Ib $\lambda\lambda\lambda$5167,5173,5184
(hereafter MgIb) lines and around CaII
$\lambda\lambda\lambda$8498,8542,8662 (hereafter CaT). 
\s~measurements were obtained from a Python-based code described in
detail in Paper I \& II. In short, it is based on the algorithm by
\citet{van94}, fitting a linear combination of Gaussian-broadened
(30--500\,km\,s$^{-1}$) template spectra (G and K
giants of various temperatures as well as spectra of A0 and F2 giants from the Indo-US survey)
and a polynomial continuum using a MCMC routine
(with the best derived $\sigma$ measurements corresponding to the MAP values).
Telluric and AGN emission lines were masked and thus excluded from
the fit.

In Paper II, \s~measurements were derived for both aperture and spatially-resolved
spectra, i.e.,~as a function of distance from the center.
The extracted spatially-resolved spectra were used to
determine the velocity dispersion within the spheroid
effective radius $\sigma_{\rm spat, reff}$, free from broadening due to a
rotating disk component. 
(Note that this assumes that the spheroid component dominates
the velocity dispersion within the spheroid effective radius,
and that contributions from bar and disk are negligible in comparison.)
To do so, we calculate the velocity dispersion within the
spheroid effective radius as determined from the surface
photometry:
\begin{eqnarray}
\sigma_{\rm spat, reff}^2 = \frac{\int\limits_{\rm -R_{eff}}^{\rm R_{eff}} (\sigma_{\rm spat}^2 (r) + v_{\rm spat}^2 (r))\cdot I (r) \cdot r \cdot dr}{\int\limits_{\rm -R_{eff}}^{\rm R_{eff}} I(r) \cdot r \cdot dr}
\end{eqnarray}
with $I(r)$ = $I (\rm {reff}) \cdot \exp (-\kappa_n \cdot [(r/r_{\rm reff})^{1/n}-1])$ 
the surface brightness of the spheroid
fitted as a S{\'e}rsic profile. Here, $v_{\rm spat}$ is the rotational component of the spheroid.
We approximated $\kappa_n$ = 1.9992$n$-0.3271
\citep[valid for 0.5 $<n<$ 10,][]{cap89,pru97}.
(Note that $n$, $r_{\rm reff}$, and $I (\rm {reff}$ are taken from the image analysis.)
Since stellar velocity dispersions were measured for spectra on both sides of the center,
along the major axis, we integrate from ``$\rm{-R_{eff}}$'' to ``$\rm{+R_{eff}}$''.

\noindent
A spline function is used to interpolate over the appropriate
radial range, since the $\sigma_{\rm spat}$ measurements are discrete.
In Section~\ref{sub:svd}, we discuss other $\sigma$ definitions used in the literature and compare them
with the fiducial value used throughout this Paper from Eq. 1.

\subsection{Black Hole Mass}
\label{subsec:mbh}

We measure the second moment of the broad H$\beta$ emission line 
from the central blue Keck/LRIS spectrum (1 square-arcsecond in size), 
using the IDL-based code implemented by \citet{par15},
allowing for a multi-component spectral decomposition of the H$\beta$ region.
Here, we briefly summarize the procedure.
First, we model and subtract the observed continuum by simultaneously
fitting the pseudo-continuum, which consists of the AGN featureless power-law
continuum, the AGN \ion{Fe}{2} emission template from \citet{bor92},
and the host-galaxy starlight templates from the Indo-US spectral library \citep{Valdes+04},
in the emission-line free windows of $4430-4770$ \AA\ and $5080-5450$ \AA.
Second, we model the continuum-subtracted H$\beta$ emission-line region
by simultaneously fitting Gauss-Hermite series 
\citep{van93,mcg08,woo06} to the broad and narrow
H$\beta$ emission lines and the [\ion{O}{3}] $\lambda\lambda4959, 5007$ narrow
emission lines and fitting Gaussian functions to the broad and narrow
\ion{He}{2} $\lambda4686$ emission lines (when blended with
the broad H$\beta$ component; see Section 3.1 of \citealt{par15} for
details). The final fits are shown in Figs.~\ref{hbeta1}-\ref{hbeta2}.

From the resulting fit, the second moment of the broad H$\beta$
component ($\sigma_{H\beta}$)
is combined with the 5100\AA~AGN luminosity derived from surface photometry
to estimate \mbh:
\begin{eqnarray}
\log M_{\rm BH} = 0.71 + 6.849 + 2 \log \left(\frac{\sigma_{\rm H_\beta}}{1000\,{\rm km\,s^{-1}}}\right)\\
+ 0.549 \log \left(\frac{\lambda L_{5100}}{10^{44}\,{\rm erg\,s^{-1}}}\right)
\end{eqnarray}
This equation is derived from adopting the most recent broad-line
region (BLR) radius-luminosity relation \citep{ben13}[Table 14,
Clean2+ExtCorr] and a virial factor of $\log f = 0.71$ 
\citep{par12,woo13}.
The results are given in Tables~\ref{table:meas1}--\ref{table:meas2}
for a sample of 79 objects (see next paragraph).  We assume a nominal
uncertainty of the BH masses measured via the virial method of 0.4 dex
\citep{ves06}. 

Note that from the full sample of 103 objects presented in Paper II,
six objects showed only a broad H$\alpha$ line in the SDSS spectrum
and no broad H$\beta$ line in either the SDSS or Keck spectra.
While we can still estimate \mbh~from the broad H$\alpha$ line in the
SDSS spectrum, we decided to exclude them from the sample,
for consistent \mbh~measurements.
An additional eight objects showed broad H$\alpha$ and H$\beta$ lines
in the SDSS spectrum, but did not reveal any broad H$\beta$ line in
the Keck spectrum. 
We excluded these objects here as well, and our \mbh~sample is thus comprised of 79 objects.
However, we will discuss them individually in an upcoming paper (V. N. Bennert et
al. 2015, in prep.). This upcoming paper will also include a direct
comparison between the SDSS spectrum and the Keck spectrum,
to study any broad-line variability.
57 objects are included in the BH mass function study by \citet{gre07}
with BH masses derived from the broad H$\alpha$ line and luminosity in the SDSS spectra,
with an overall good agreement in the mass measurements.

\section{FINAL SAMPLE AND COMPARISON SAMPLES}
\label{sec:final}
While a sample of 103 objects was observed at Keck, not all properties
could be determined for all objects (see Section~\ref{sec:derived}).
Taking into account the overlap between the measurements of surface
photometry, stellar velocity dispersion and \mbh, our final sample for
the \mbh-\s~relation consists of 66 objects (see
Tables~\ref{table:meas1}--\ref{table:meas2}).

We compare our sample with the compilation of
\mbh~and \s~for quiescent galaxies \citep[][72 objects]{mcc13}
as well as reverberation-mapped AGNs \citep[][29 objects; adopting the same virial factor as
for our sample; $\log f = 0.71$]{woo15}.
In Table~\ref{fits_relations}, we additionally compare with the recent compilation of
\mbh~and \s~for quiescent galaxies by \citet{kor13}
(51 objects; pseudo bulges and mergers excluded).
Our results do not change depending on what comparison sample we use.
We discuss below the effects of different $\sigma$ definitions used in the literature,
including these comparison samples. Unfortunately, none of the literature uses the definition
that we consider the most robust in this paper (see discussion below).

Note that while for the quiescent galaxies, BH masses have been
derived from direct dynamical measurements,
the BH masses for active galaxies are calibrated
masses either from reverberation mapping or from
the virial method.

\begin{deluxetable*}{lccccccccccc}
\tabletypesize{\scriptsize}
\tablecolumns{12}
\tablecaption{Sample and Derived Quantities}
\tablehead{
\colhead{Object} & 
\colhead{R.A.} &
\colhead{Decl.} &
\colhead{$z$} &
\colhead{$\sigma_{\rm spat, reff}$} &
\colhead{$r_{\rm eff, sph}$} &
\colhead{$\sigma_{H\beta}$} & 
\colhead{$\lambda L_{5100}$} & 
\colhead{log \mbh/\msun} &
\colhead{Host} &
\colhead{Spheroid} &
\colhead{Alt. Name} \\
& \colhead{(J2000)} & \colhead{(J2000)} & & \colhead{(km\,s$^{-1}$)} 
& \colhead{('')}
& \colhead{(km\,s$^{-1}$)} 
& \colhead{($10^{44}$\,erg\,s$^{-1}$)} 
& & & \\
\colhead{(1)} & \colhead{(2)}  & \colhead{(3)} & \colhead{(4)} &
\colhead{(5)} & \colhead{(6)} & \colhead{(7)}
& \colhead{(8)} & \colhead{(9)} & \colhead{(10)} & \colhead{(11)} & \colhead{(12)}}
\startdata
0013$$-$$0951 & 00 13 35.38 & $-$09  51  20.9 & 0.0615 &   96 & 4.00 & 2111$\pm$211 & 0.225 & 7.85 & BD & C &  \\ 
0026+0009 & 00 26 21.29 & +00  09  14.9 & 0.0600 &  172 & 1.54 & 1527$\pm$227 & 0.025 & 7.05 & BDB & C &  \\ 
0038+0034 & 00 38 47.96 & +00  34  57.5 & 0.0805 &  127 & 1.23 & 3328$\pm$239 & 0.208 & 8.23 & BD & C &  \\ 
0109+0059 & 01 09 39.01 & +00  59  50.4 & 0.0928 &  183 & 0.20 & 1797$\pm$268 & 0.101 & 7.52 & BD & C &  \\ 
0121$$-$$0102 & 01 21 59.81 & $-$01  02  24.4 & 0.0540 &   90 & 1.74 & 1742$\pm$106 & 0.290 & 7.75 & BDB & P & MRK1503 \\ 
0150+0057 & 01 50 16.43 & +00  57  01.9 & 0.0847 &  176 & 2.85 & 2057$\pm$129 & 0.020 & 7.25 & BD & C &  \\ 
0206$$-$$0017 & 02 06 15.98 & $-$00  17  29.1 & 0.0430 &  225 & 7.29 & 1979$\pm$185 & 0.540 & 8.00 & BD(M) & C & UGC1597 \\ 
0212+1406 & 02 12 57.59 & +14  06  10.0 & 0.0618 &  171 & 0.83 & 1586$\pm$86 & 0.069 & 7.32 & BD & C &  \\ 
0301+0115 & 03 01 44.19 & +01  15  30.8 & 0.0747 &   99 & 1.90 & 1653$\pm$105 & 0.155 & 7.55 & B & C &  \\ 
0310$$-$$0049 & 03 10 27.82 & $-$00  49  50.7 & 0.0801 & ... & 0.20 & 1558$\pm$334 & 1.172 & 7.98 & BD & C &  \\ 
0336$$-$$0706 & 03 36 02.09 & $-$07  06  17.1 & 0.0970 &  236 & 7.17 & 2403$\pm$192 & 0.036 & 7.53 & BD & C &  \\ 
0353$$-$$0623 & 03 53 01.02 & $-$06  23  26.3 & 0.0760 &  175 & 1.11 & 1548$\pm$537 & 0.160 & 7.50 & BD & C &  \\ 
0731+4522 & 07 31 26.68 & +45  22  17.4 & 0.0921 & ... & 1.39 & 1885$\pm$134 & 0.089 & 7.53 & BD & C &  \\ 
0737+4244 & 07 37 03.28 & +42  44  14.6 & 0.0882 & ... & 2.53 & 1692$\pm$98 & 0.141 & 7.55 & BD & C &  \\ 
0802+3104 & 08 02 43.40 & +31  04  03.3 & 0.0409 &  116 & 3.41 & 1772$\pm$185 & 0.072 & 7.43 & BD & C &  \\ 
0811+1739 & 08 11 10.28 & +17  39  43.9 & 0.0649 &  142 & 1.98 & 1520$\pm$361 & 0.042 & 7.17 & BD & C &  \\ 
0813+4608 & 08 13 19.34 & +46  08  49.5 & 0.0540 &  122 & 0.99 & 1430$\pm$91 & 0.048 & 7.14 & BDB & P &  \\ 
0845+3409 & 08 45 56.67 & +34  09  36.3 & 0.0655 &  123 & 1.15 & 1718$\pm$172 & 0.064 & 7.37 & BDB & P & KUG0842+343A \\ 
0854+1741 & 08 54 39.25 & +17  41  22.5 & 0.0654 & ... & 1.98 & 1472$\pm$269 & 0.270 & 7.58 & BD & C & MRK1220 \\ 
0857+0528 & 08 57 37.77 & +05  28  21.3 & 0.0586 &  126 & 2.22 & 1485$\pm$51 & 0.135 & 7.42 & BD & C &  \\ 
0904+5536 & 09 04 36.95 & +55  36  02.5 & 0.0371 &  194 & 5.43 & 2483$\pm$36 & 0.088 & 7.77 & B(M) & C &  \\ 
0921+1017 & 09 21 15.55 & +10  17  40.9 & 0.0392 &   83 & 3.37 & 2317$\pm$286 & 0.030 & 7.45 & BD & C & VIIIZw045 \\ 
0923+2254 & 09 23 43.00 & +22  54  32.7 & 0.0332 &  149 & 1.43 & 1824$\pm$265 & 0.194 & 7.69 & BDB & P &  \\ 
0923+2946 & 09 23 19.73 & +29  46  09.1 & 0.0625 &  142 & 3.52 & 2936$\pm$247 & 0.019 & 7.56 & B & C &  \\ 
0927+2301 & 09 27 18.51 & +23  01  12.3 & 0.0262 &  196 & 13.42 & 2112$\pm$205 & 0.005 & 6.94 & BD & C & NGC2885 \\ 
0932+0233 & 09 32 40.55 & +02  33  32.6 & 0.0567 &  126 & 0.63 & 1814$\pm$72 & 0.069 & 7.44 & BD & C &  \\ 
0936+1014 & 09 36 41.08 & +10  14  15.7 & 0.0600 & ... & 3.25 & 1995$\pm$80 & 0.091 & 7.59 & BD & C &  \\ 
1029+1408 & 10 29 25.73 & +14  08  23.2 & 0.0608 &  185 & 2.57 & 2456$\pm$344 & 0.133 & 7.86 & BDB & C &  \\ 
1029+2728 & 10 29 01.63 & +27  28  51.2 & 0.0377 &  112 & 3.41 & 1544$\pm$252 & 0.014 & 6.92 & B & C &  \\ 
1029+4019 & 10 29 46.80 & +40  19  13.8 & 0.0672 &  166 & 1.54 & 2193$\pm$387 & 0.093 & 7.68 & BD & C &  \\ 
1042+0414 & 10 42 52.94 & +04  14  41.1 & 0.0524 &   74 & 3.13 & 1518$\pm$102 & 0.039 & 7.14 & B & C &  \\ 
1043+1105 & 10 43 26.47 & +11  05  24.3 & 0.0475 & ... & 3.09 & 2314$\pm$28 & 0.171 & 7.87 & B & C &  \\ 
1049+2451 & 10 49 25.39 & +24  51  23.7 & 0.0550 &  162 & 1.23 & 2534$\pm$135 & 0.246 & 8.03 & BD & C &  \\ 
1058+5259 & 10 58 28.76 & +52  59  29.0 & 0.0676 &  122 & 0.99 & 1896$\pm$645 & 0.075 & 7.50 & BDB & P &  \\ 
1101+1102 & 11 01 01.78 & +11  02  48.8 & 0.0355 &  197 & 8.16 & 3949$\pm$170 & 0.068 & 8.11 & BD & C & MRK728 \\ 
1104+4334 & 11 04 56.03 & +43  34  09.1 & 0.0493 &   87 & 1.15 & 1719$\pm$160 & 0.016 & 7.04 & BD & C &  \\ 
1116+4123 & 11 16 07.65 & +41  23  53.2 & 0.0210 &  108 & 3.76 & 3136$\pm$384 & 0.004 & 7.23 & BD & C & UGC6285 \\ 
1132+1017 & 11 32 49.28 & +10  17  47.4 & 0.0440 & ... & 1.94 & 1900$\pm$86 & 0.049 & 7.40 & BDB & C & IC2921 \\ 
1137+4826 & 11 37 04.17 & +48  26  59.2 & 0.0541 &  155 & 1.07 & 1606$\pm$92 & 0.006 & 6.74 & B & C &  \\ 
1143+5941 & 11 43 44.30 & +59  41  12.4 & 0.0629 &  122 & 3.13 & 1790$\pm$128 & 0.099 & 7.51 & BD & C &  \\ 
1144+3653 & 11 44 29.88 & +36  53  08.5 & 0.0380 &  168 & 1.39 & 2933$\pm$205 & 0.041 & 7.73 & BDB & P & KUG1141+371 \\ 
1145+5547 & 11 45 45.18 & +55  47  59.6 & 0.0534 &  118 & 1.31 & 1837$\pm$208 & 0.027 & 7.22 & BDB & P &  \\ 
1147+0902 & 11 47 55.08 & +09  02  28.8 & 0.0688 &  147 & 2.61 & 2896$\pm$188 & 0.690 & 8.39 & B & C &  \\ 
1205+4959 & 12 05 56.01 & +49  59  56.4 & 0.0630 &  152 & 2.02 & 2678$\pm$294 & 0.177 & 8.00 & BD & C &  \\ 
1210+3820 & 12 10 44.27 & +38  20  10.3 & 0.0229 &  141 & 1.23 & 2831$\pm$148 & 0.062 & 7.80 & BD & C & KUG1208+3806 \\ 
1216+5049 & 12 16 07.09 & +50  49  30.0 & 0.0308 &  189 & 6.57 & 4487$\pm$477 & 0.035 & 8.06 & BDB & C & MRK1469 \\ 
1223+0240 & 12 23 24.14 & +02  40  44.4 & 0.0235 &  124 & 7.25 & 2306$\pm$107 & 0.007 & 7.10 & B & C & MRK50 \\ 
1231+4504 & 12 31 52.04 & +45  04  42.9 & 0.0621 &  169 & 1.23 & 1555$\pm$168 & 0.073 & 7.32 & BD(M) & C &  \\ 
1241+3722 & 12 41 29.42 & +37  22  01.9 & 0.0633 &  144 & 1.43 & 1574$\pm$100 & 0.091 & 7.38 & BD & C &  \\ 
1246+5134 & 12 46 38.74 & +51  34  55.9 & 0.0668 &  119 & 3.05 & 1141$\pm$130 & 0.044 & 6.93 & BD & C &  \\ 
1306+4552 & 13 06 19.83 & +45  52  24.2 & 0.0507 &  114 & 2.34 & 1892$\pm$297 & 0.018 & 7.16 & BD & C &  \\ 
1307+0952 & 13 07 21.93 & +09  52  09.3 & 0.0490 & ... & 3.21 & 1630$\pm$165 & 0.041 & 7.22 & BD & C &  \\ 
1312+2628 & 13 12 59.59 & +26  28  24.0 & 0.0604 &  109 & 1.47 & 1572$\pm$496 & 0.154 & 7.51 & BD & C &  \\ 
1323+2701 & 13 23 10.39 & +27  01  40.4 & 0.0559 &  124 & 0.87 & 2414$\pm$376 & 0.026 & 7.45 & BD & C &  \\ 
1355+3834 & 13 55 53.52 & +38  34  28.5 & 0.0501 & ... & 2.77 & 4034$\pm$301 & 0.097 & 8.21 & B & C & MRK464 \\ 
1405$$-$$0259 & 14 05 14.86 & $-$02  59  01.2 & 0.0541 &  125 & 0.59 & 1599$\pm$140 & 0.020 & 7.04 & BD & C &  \\ 
1416+0137 & 14 16 30.82 & +01  37  07.9 & 0.0538 &  173 & 3.41 & 1514$\pm$233 & 0.064 & 7.26 & BD & C &  \\ 
1419+0754 & 14 19 08.30 & +07  54  49.6 & 0.0558 &  215 & 4.99 & 3006$\pm$371 & 0.116 & 8.00 & BD & C &  \\ 
1434+4839 & 14 34 52.45 & +48  39  42.8 & 0.0365 &  109 & 1.23 & 1731$\pm$85 & 0.210 & 7.66 & BDB & C & NGC5683 \\ 
1505+0342 & 15 05 56.55 & +03  42  26.3 & 0.0358 & ... & 2.10 & 2127$\pm$139 & 0.382 & 7.98 & BD & C & MRK1392 \\ 
1535+5754 & 15 35 52.40 & +57  54  09.3 & 0.0304 &  110 & 4.51 & 2442$\pm$93 & 0.287 & 8.04 & B & C & MRK290 \\ 
1543+3631 & 15 43 51.49 & +36  31  36.7 & 0.0672 &  146 & 2.93 & 1820$\pm$168 & 0.229 & 7.73 & BD & C &  \\ 
1545+1709 & 15 45 07.53 & +17  09  51.1 & 0.0481 &  163 & 1.15 & 3588$\pm$226 & 0.070 & 8.03 & BD & C &  \\ 
1554+3238 & 15 54 17.42 & +32  38  37.6 & 0.0483 &  158 & 1.82 & 2523$\pm$159 & 0.125 & 7.87 & BD & C &  \\ 
1557+0830 & 15 57 33.13 & +08  30  42.9 & 0.0465 & ... & 1.62 & 2388$\pm$91 & 0.063 & 7.66 & B & C &  \\ 
\enddata
  \tablecomments{
Col. (1): target ID used throughout the text (based on R.A. and declination). 		      	   	  
Col. (2): right ascension. 				  
Col. (3): declination. 
Col. (4): redshift from SDSS-DR7.	
Col. (5): spatially resolved stellar-velocity dispersion within
spheroid effective radius. Determined from CaT, if available,
else MgIb or CaHK (Paper II) according to Equation 1 (uncertainty of 0.04 dex).
Col. (6): spheroid effective radius in arcseconds.
Col. (7): second moment of broad H$\beta$.			  
Col. (8): rest-frame luminosity at 5100\AA~determined from SDSS g' band surface photometry (fiducial error 0.1 dex).
Col. (9): logarithm of BH mass (solar units) (uncertainty of 0.4 dex).
Col. (10): Host-galaxy decomposition: B = ``bulge only'', BD =
``bulge+disk'', BDB = ``bulge+disk+bar''. The ``(M)'' indicates an
interacting or merging galaxy.
Col. (11): Classification of spheroid as either classical bulge (C) or pseudo bulge (P).
Col. (12): Alternative name from NED.}
\label{table:meas1}
\end{deluxetable*}

\begin{deluxetable*}{lccccccccccc}
\tabletypesize{\scriptsize}
\tablecolumns{12}
\tablecaption{Sample and Derived Quantities}
\tablehead{
\colhead{Object} & 
\colhead{R.A.} &
\colhead{Decl.} &
\colhead{$z$} &
\colhead{$\sigma_{\rm spat, reff}$} &
\colhead{$r_{\rm eff, sph}$} &
\colhead{$\sigma_{H\beta}$} & 
\colhead{$\lambda L_{5100}$} & 
\colhead{log \mbh/\msun} &
\colhead{Host} &
\colhead{Spheroid} &
\colhead{Alt. Name} \\
& \colhead{(J2000)} & \colhead{(J2000)} & & \colhead{(km\,s$^{-1}$)} 
& \colhead{('')}
& \colhead{(km\,s$^{-1}$)} 
& \colhead{($10^{44}$\,erg\,s$^{-1}$)} 
& & & \\
\colhead{(1)} & \colhead{(2)}  & \colhead{(3)} & \colhead{(4)} &
\colhead{(5)} & \colhead{(6)} & \colhead{(7)}
& \colhead{(8)} & \colhead{(9)} & \colhead{(10)} & \colhead{(11)} & \colhead{(12)}}
\startdata
1605+3305 & 16 05 02.46 & +33  05  44.8 & 0.0532 &  187 & 1.58 & 1960$\pm$272 & 0.254 & 7.82 & B & C &  \\ 
1606+3324 & 16 06 55.94 & +33  24  00.3 & 0.0585 &  157 & 1.54 & 2053$\pm$80 & 0.067 & 7.54 & BD & C &  \\ 
1611+5211 & 16 11 56.30 & +52  11  16.8 & 0.0409 &  116 & 1.66 & 2515$\pm$410 & 0.056 & 7.67 & BD & C &  \\ 
1636+4202 & 16 36 31.28 & +42  02  42.5 & 0.0610 &  205 & 8.24 & 2492$\pm$230 & 0.125 & 7.86 & BD & P &  \\ 
1708+2153 & 17 08 59.15 & +21  53  08.1 & 0.0722 &  231 & 5.86 & 2975$\pm$122 & 0.276 & 8.20 & B(M) & C &  \\ 
2116+1102 & 21 16 46.33 & +11  02  37.3 & 0.0805 & ... & 10.38 & 2484$\pm$42 & 0.220 & 7.99 & BD & C &  \\ 
2140+0025 & 21 40 54.55 & +00  25  38.2 & 0.0838 &  126 & 1.90 & 1114$\pm$64 & 0.585 & 7.52 & B & C &  \\ 
2215$$-$$0036 & 22 15 42.29 & $-$00  36  09.6 & 0.0992 & ... & 5.66 & 1636$\pm$92 & 0.202 & 7.61 & BD & C &  \\ 
2221$$-$$0906 & 22 21 10.83 & $-$09  06  22.0 & 0.0912 &  142 & 3.60 & 2375$\pm$131 & 0.104 & 7.77 & B & C &  \\ 
2222$$-$$0819 & 22 22 46.61 & $-$08  19  43.9 & 0.0821 &  122 & 1.07 & 1799$\pm$168 & 0.177 & 7.66 & BD & C &  \\ 
2233+1312 & 22 33 38.42 & +13  12  43.5 & 0.0934 &  193 & 1.19 & 2477$\pm$135 & 0.368 & 8.11 & BD & C &  \\ 
2254+0046 & 22 54 52.24 & +00  46  31.4 & 0.0907 & ... & 2.73 &  989$\pm$261 & 0.481 & 7.37 & B(M) & C &  \\ 
2327+1524 & 23 27 21.97 & +15  24  37.4 & 0.0458 &  225 & 7.29 & 1924$\pm$166 & 0.079 & 7.52 & B & C &  \\ 
2351+1552 & 23 51 28.75 & +15  52  59.1 & 0.0963 &  186 & 1.43 & 2974$\pm$144 & 0.165 & 8.08 & B & C &  \\ 
\enddata
\tablecomments{Table~\ref{table:meas1} continued.
                 }
\label{table:meas2}
\end{deluxetable*}

 \newpage
\section{RESULTS AND DISCUSSION}
\label{sec:results}
For the discussion of our results, we only consider our final sample of 66 objects
for which we have both \mbh~and \s~measurements.

\subsection{Host-galaxy Morphologies}
\label{subsec:morph}
We visually inspected the multi-filter SDSS images
as well as the fits and residuals
to determine the best host-galaxy decomposition.
The majority of the host galaxies are classified as Sa or later
(51/66 = 77\%) and was fitted either by a spheroid+disk decomposition
(40) or spheroid+disk+bar (11).
For the remaining 15 objects, a spheroid only fit was deemed sufficient.
The high fraction of spiral galaxies is typical for a sample of
(mostly radio-quiet) Seyfert galaxies \citep[e.g.,][and references therein]{hun99}.
This is consistent with the majority of objects ($\sim$60\%) showing
rotation curves with a maximum velocity between 100 and 200
km\,s$^{-1}$.
6\% of the sample (4/66) are merging or interacting galaxies.
This is lower than for our high-redshift Seyfert galaxies
\citep[$\sim$30\% at $z \simeq 0.4-0.6$][]{par15}
and more comparable to quiescent galaxies in the local Universe.

Of the 79 objects in our sample, 75 are included in the morphological classification by Galaxy Zoo \citep{lin11},
but for only 28 did the vote reach the necessary 80\% mark to flag the morphology as either spiral or elliptical.
For those, our classification agrees in the majority of cases (82\%) with the rest being classified as ellipticals by
\citet{lin11} while we classified them as spirals. However, 
we consider our classification as more robust since it also takes into accounts the fits and residuals,
especially given the AGN central point source, 
which could lead to an overestimation of the presence of a bulge by Galaxy Zoo.

For the spiral galaxies in our sample, 
a little more than half of spheroid components are fitted by a S{\'e}rsic
index $<2$ (28 objects = 55\%, corresponding to 43\% of the total sample).
For the rest (23 objects), the spheroid component
is fitted by a S{\'e}rsic index $\ge2$.
On average, the S{\'e}rsic index for the spiral galaxies is 2.2$\pm$1.6,
for $n\ge2$ bulges 3.8$\pm$0.8, and for $n<2$ bulges 0.9$\pm$0.3.

If we take solely the S{\'e}rsic index as an indicator for the existence of a classical 
vs. pseudo bulge, half of our spiral galaxies have pseudo bulges; we consider them candidate
pseudo bulges.
The host-galaxies of the reverberation-mapped AGNs have a similar distribution:
\citet{ho14} classify 75\% as spiral galaxies with roughly half having classical bulges and half having pseudo bulges.
In a bulge+disk decomposition of galaxies in SDSS, 
roughly 25\% of galaxies with sufficient 
image quality to study their bulge profile shape (a total of $\sim$53,000) have pseudo bulges,
if we use the same criterion with S{\'e}rsic index $<2$
\citep[][their Fig. 15, when excluding bars with $n=0.5$]{sim11}.

However, we further follow the guidelines by \citet{kor13} to distinguish
between classical bulges and pseudo bulges,
applying the following four criteria that we can probe with our data:
(i) S{\'e}rsic index $n<2$ for pseudo bulges, $n\ge2$ for classical bulges;
(ii) bulge-to-total luminosity ratios (in all four SDSS bands) 
B/T > 0.5 for classical bulges;
(iii) $v_{\rm max, reff}$/$\sigma_{\rm center}$ $>$ 1 for pseudo bulges, < 1 for classical bulges
(here, $v_{\rm max, reff}$ is the maximum velocity at the effective radius of the bulge 
and $\sigma_{\rm center}$ is the stellar velocity dispersion in the center);
(iv) the presence of a bar in face-on galaxies as an indicator of a pseudo bulge.
To be conservative, we only classify objects as having a pseudo bulge for which at least three of the above four criteria are met.
That leaves us with a total of eight spiral galaxies with a definite pseudo bulge.

\subsection{Stellar-velocity Dispersions}
\label{sub:svd}
The measurement of the stellar velocity dispersion profiles and
rotation curves is described in detail in Paper II. In this paper,
with the addition of the surface photometry parameters, we have all
the necessary information to investigate the systematic uncertainties
and biases related to the definition of the stellar velocity dispersion.

With this goal in mind, we carry out a systematic comparison between
different definitions of the velocity dispersion parameter taken from
the literature and our fiducial measurement ($\sigma_{\rm spat,
reff}$, Eq. 1). Specifically we compute velocity dispersions as
average of the second moment of the velocity field, by varying the
size of the aperture, by considering the difference between
correcting and not correcting the velocity rotation curve for
inclination, and by considering the effects of contamination by nuclear
light. While not 100\% exhaustive, our list of definitions includes
most of the choices adopted in the literature. For example,
\citet{fer00} use the second moment, integrating over one quarter of the effective radius of the galaxy
and correcting the velocity for inclination.  Neither \citet{gue09}, \citet{mcc13}, nor
\citet{kor13} correct the velocity for inclination. However,
\citet{gue09} and \citet{mcc13} use the effective radius of the galaxy, while \citet{kor13} use half of the effective radius of the galaxy.
\citet{mcc13} additionally discuss the choice of the minimum radius and find that setting it to zero
\citep[as done by e.g.,][]{gue09} can result in $\sigma$ values smaller by 10-15\%,
since it includes signal from within the BH gravitational sphere of influence. Thus, they instead
set the minimum radius to the latter. However, for our data, we are not resolving the BH gravitational sphere of influence.
Finally, many $\sigma$ measurements in the literature are derived from aperture spectra,
such as SDSS fiber spectra or spectra of distant galaxies
integrated over different aperture sizes, making a direct comparison difficult. 
The reverberation-mapped AGN comparison sample
falls into this category \citep{woo15}. 

\begin{figure}
\includegraphics[width=\linewidth]{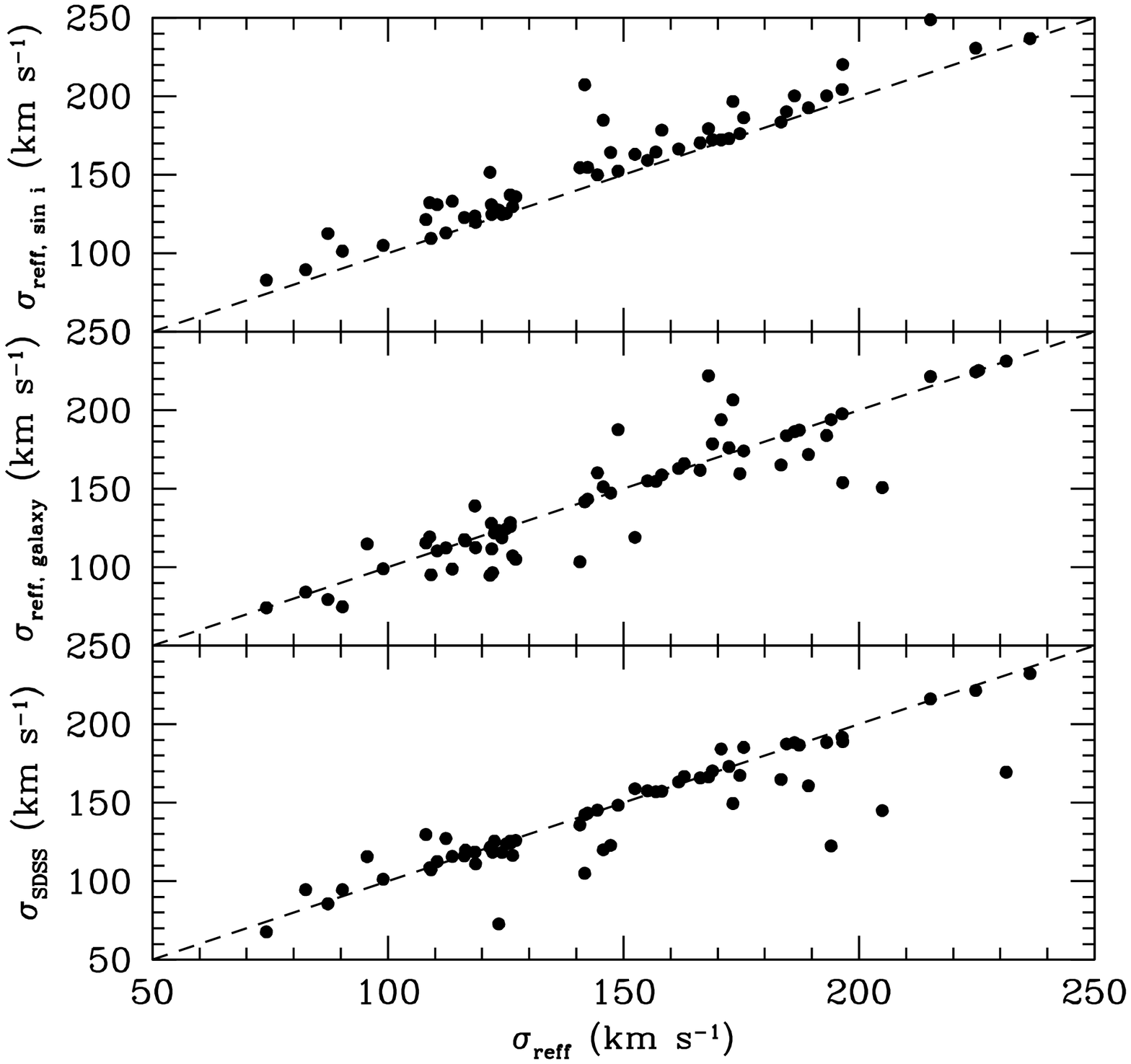}
\caption{Comparison between stellar velocity dispersion as measured according to different definitions. 
The fiducial definition used throughout this paper,
$\sigma_{\rm spat, reff}$, (Eq. 1) is shown on the x-axis, 
while alternative definitions based on different apertures (lower panel: 'SDSS' corresponds to $1\farcs5$ radius; middle panel: 'galaxy' corresponds to the effective radius of the entire galaxy) or including corrections to the velocity for inclination (upper panel: 'sini'). The dashed line represents the identity.}
\label{fig_sigmacompare}
\end{figure}

Table~\ref{table_sigmacompare} lists the results for different
possible comparisons, and Fig.~\ref{fig_sigmacompare} shows three
examples.  Here, $\sigma_{\rm spat, 0.5 reff}$ ($\sigma_{\rm spat, 0.25 reff}$) integrates out to half (one quarter) of the effective
bulge radius.  $\sigma_{\rm spat, reff, sini}$ additionally corrects the velocity for inclination, as estimated from the
disk:
\begin{eqnarray}
\sigma_{\rm spat, reff, sini}^2 = \frac{\int\limits_{\rm -R_{eff}}^{\rm R_{eff}} (\sigma_{\rm spat}^2 (r) + v_{\rm spat}^2/\sin(i)^2 (r))\cdot I (r) \cdot r \cdot dr}{\int\limits_{\rm -R_{eff}}^{\rm R_{eff}} I(r) \cdot r \cdot dr}
\end{eqnarray}
Again, we adopt smaller integration radii in
$\sigma_{\rm spat, 0.5 reff, sini}$ and 
$\sigma_{\rm spat, 0.25 reff, sini}$ (half and one quarter of the effective bulge radius, respectively).
$\sigma_{\rm spat, reff, galaxy}$ integrates out to the effective radius of the galaxy instead,
or alternatively half/one quarter of that ($\sigma_{\rm spat, 0.5 reff,galaxy}$; $\sigma_{\rm spat, 0.25 reff,galaxy}$).
It follows that $\sigma_{\rm spat, reff, galaxy, sini}$ corrects the velocity for inclination,
considering different integration limits in
$\sigma_{\rm spat, 0.5 reff, galaxy, sini}$ and $\sigma_{\rm spat, 0.25 reff, galaxy, sini}$.
Finally, $\sigma_{\rm spat, SDSS}$ 
integrates within the 1.5'' radius of the SDSS fiber.
(Note that in fact our $\sigma_{\rm ap, SDSS}$ corresponds to a
rectangular region with 1\farcs5 radius and 1$\arcsec$~width, given
the width of the long slit used.)

\newpage
The largest (average) effect on the derived $\sigma$ is the correction of the
velocity for inclination, which can result in $\sigma$ measurements by
on average 31$\pm$9\% larger (see Fig.~\ref{fig_sigmacompare}) 
(43$\pm$9\% in case of the galaxy effective radius), with
individual objects as much as doubled.  This discrepancy is reduced by
considering smaller integration limits, since rotation is negligible
at the center of most objects.
However, such large differences are mainly due to galaxies seen close
to face on, for which the velocities are highly uncertain and
sometimes inflated beyond reasonably physical limits, resulting in
large outliers. For example, if we impose arbitrarily that
$v/\sin(i)<\sigma$ (i.e., rotation not dominant), the average ratio
falls to 1.06$\pm$0.01.
Or if instead less stringent we impose that $v/\sin(i)< 2 \sigma$ 
(e.g.,~pseudo bulges may be rotation dominated), the average ratio is
1.14$\pm$0.02.
Based on this source of uncertainty, we
emphasize that the large difference should not be taken at face value,
but more as an indication that for face-on galaxies we simply do not
know the contribution of the rotational support. The intrinsic
uncertainty in inclination correction for face-on objects is
compounded by the fact that we do not have a good estimate of the
inclination of the bulge. In the definitions of $\sigma$ above labeled
as 'sini', we used the inclination of the disk component as a proxy.
Given these caveats, we decided to not correct the velocities for
inclination in our fiducial measurement, $\sigma_{\rm spat, reff}$
(Eq. 1), which is a common practice in the literature,
including the comparison samples considered here.
However, we caution that \s\, might be underestimated in
some cases.

When increasing the radius to the galaxy effective radius,
the $\sigma$ measurement decreases on average by 1$\pm$1\%
(see Fig.~\ref{fig_sigmacompare}).
This is expected, since most objects show a decreasing
stellar velocity dispersion with distance from the center (Paper II).

However, there is another effect that
might counterbalance this trend: choosing a larger radius also
includes more and more rotational velocities, in particular those of
the disk.  Depending on the viewing angle, disk rotation can lead to
over-estimating \s~, if seen edge-on, or under-estimating \s, if seen
face-on, given that the disk is kinematically cold
\citep[e.g.,][Paper I \& II]{woo06,woo13}.
This effect can be potentially important, given the variety of host-galaxy morphologies
in both local \citep[e.g.,][]{mal98,hun04,kim08,ben09,ben11a}
and distant AGNs \citep[][]{ben10,ben11b,par15}.
When considering the different morphologies and inclinations in our
comparison between $\sigma_{\rm spat, reff}$ and $\sigma_{\rm spat,
reff,galaxy}$, we confirm the expected trend. 

\citet{bel14}  use
cosmological smoothed particle hydrodynamics (SPH) simulations of five disk galaxies and
find that the line-of-sight effect due to galaxy orientation can affect
$\sigma$ by 30\% with face-on views resulting in systematically lower velocity dispersion measurements
and edge-on orientations leading to higher values
due to a contamination of rotating disk stars.
Both fiber-based SDSS
data \citep[e.g.,][]{gre06,shen08} as well as aperture spectra for
distant galaxies \citep[e.g., our studies on the evolution of the
\mbh-\s~relation;][]{woo06,tre07,woo08} can suffer from the added
uncertainty of the effect of the disk on \s.
The evolutionary studies rely on active galaxies, by necessity,
and given the distance of the galaxies, stellar velocity dispersion measurements are challenging.
Aperture spectra typically 
include the central few kpc of the galaxy (for a typical aperture of 1\arcsec=5-7 kpc 
for the redshift range of $z=0.36-0.57$ covered by e.g., \citealt{woo08}),
and can be contaminated by the disk kinematics,
especially for Seyfert-1 galaxies chosen for these studies due to their relatively weak AGN power-law continuum.
However, the evolutionary trend found by these studies is that of
distant spheroids having on average smaller velocity dispersions than local ones.
This effect cannot be explained by disk kinematics,
since we would expect the opposite from rotational support.
Thus, the offset of the distant AGNs from the local \mbh-$\sigma$ scaling relations
is, if anything, underestimated.

Choosing a generic radius of 1$\farcs$5 as the SDSS fiber results in
an underestimation of $\sigma$ by 3$\pm$1\% (Fig.~\ref{fig_sigmacompare}).  
Note, however, that this difference is very sensitive
to the distance of the particular objects, and the ratio between aperture size
to bulge (or galaxy) effective radius. For our sample,
the effective bulge radius (on average 2.9$\pm$0.3)
is close to the SDSS fiber size, so the difference is small.

In all the $\sigma$ definitions mentioned above, we measured $\sigma$
(and the velocity) as a function of distance from the center and
integrated over it later out to the effective bulge radius (according
to Eq. 1).  However, this is not the same as directly integrating over
the entire spectrum (out to the effective radius) and then measuring
$\sigma$, since the latter will always include AGN power law and emission lines,
while in the former approach the central spectrum is sometimes
excluded due to AGN contamination (Paper II).  Thus, we
also compare our fiducial stellar velocity dispersion $\sigma_{\rm
spat, reff}$ with the corresponding value derived from aperture
spectra which on average overestimates $\sigma$ by 1$\pm$2\%.

To conclude, these comparisons show that the choice of the exact
definition of $\sigma$ can have a non-negligible effect (up to 40\%) and care needs
to be taken, especially when comparing to other values in the literature.
It is important to understand the effects of inclination, the
rotational contribution of the disk, and the AGN contribution
in the center, on the measurement of stellar velocity dispersion.
We consider the value used in this paper, 
 $\sigma_{\rm spat, reff}$ (Eq. 1), the stellar velocity dispersion within the
bulge effective radius derived from spatially resolved $\sigma$ and velocity
measurements, the most robust measurement
since it excludes contributions of disk rotation and AGN emission.
Moreover, distinguishing between bulge and disk in the context of the BH mass
scaling relations is especially important if BHs correlate only with the bulge component
and not the disk, as suggested by recent studies \citep[][]{kor11,kor13}.

\subsection{\mbh-\s~Relation}
\begin{figure*}[ht!]
\includegraphics[scale=0.42]{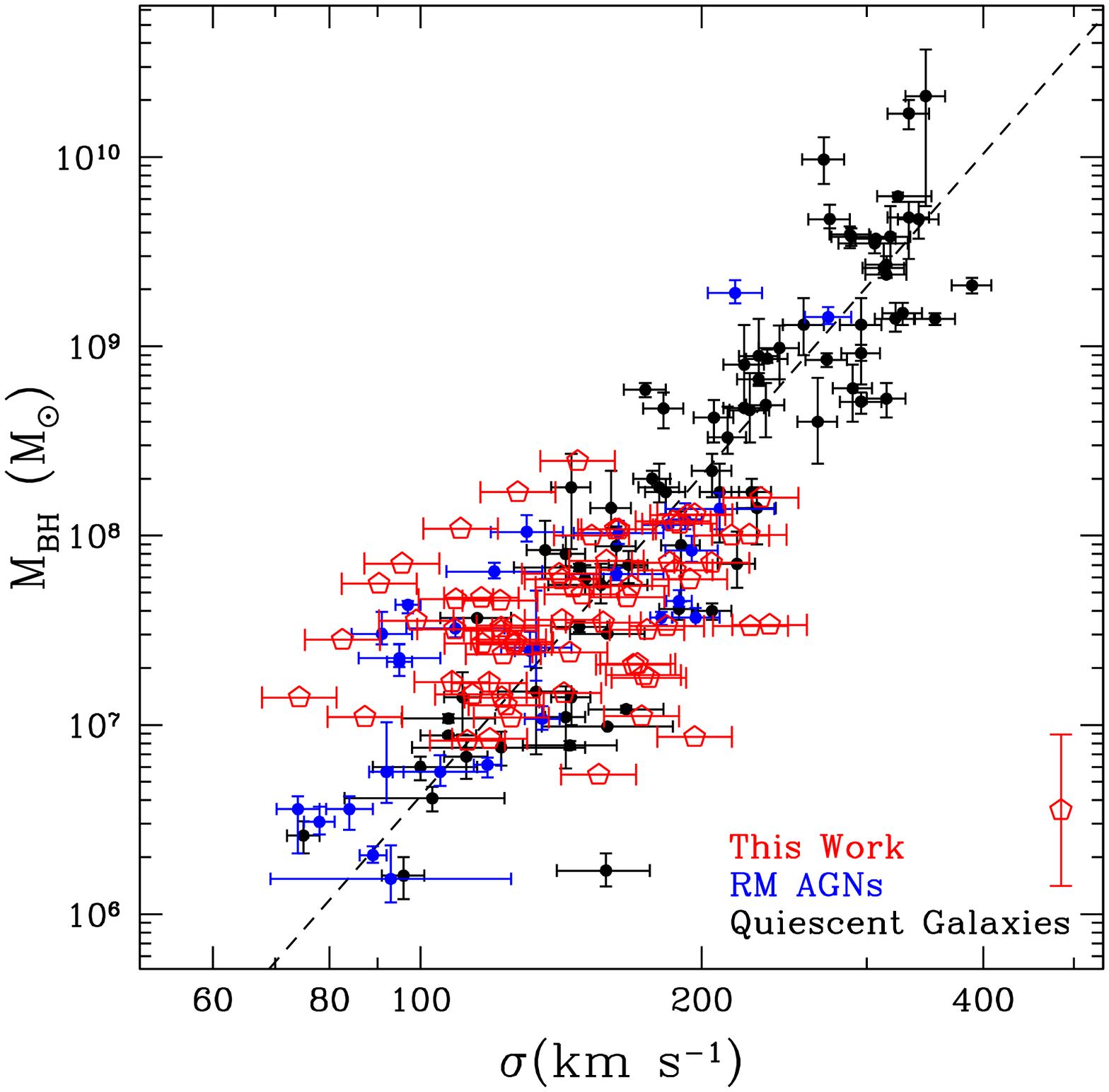}
\includegraphics[scale=0.42]{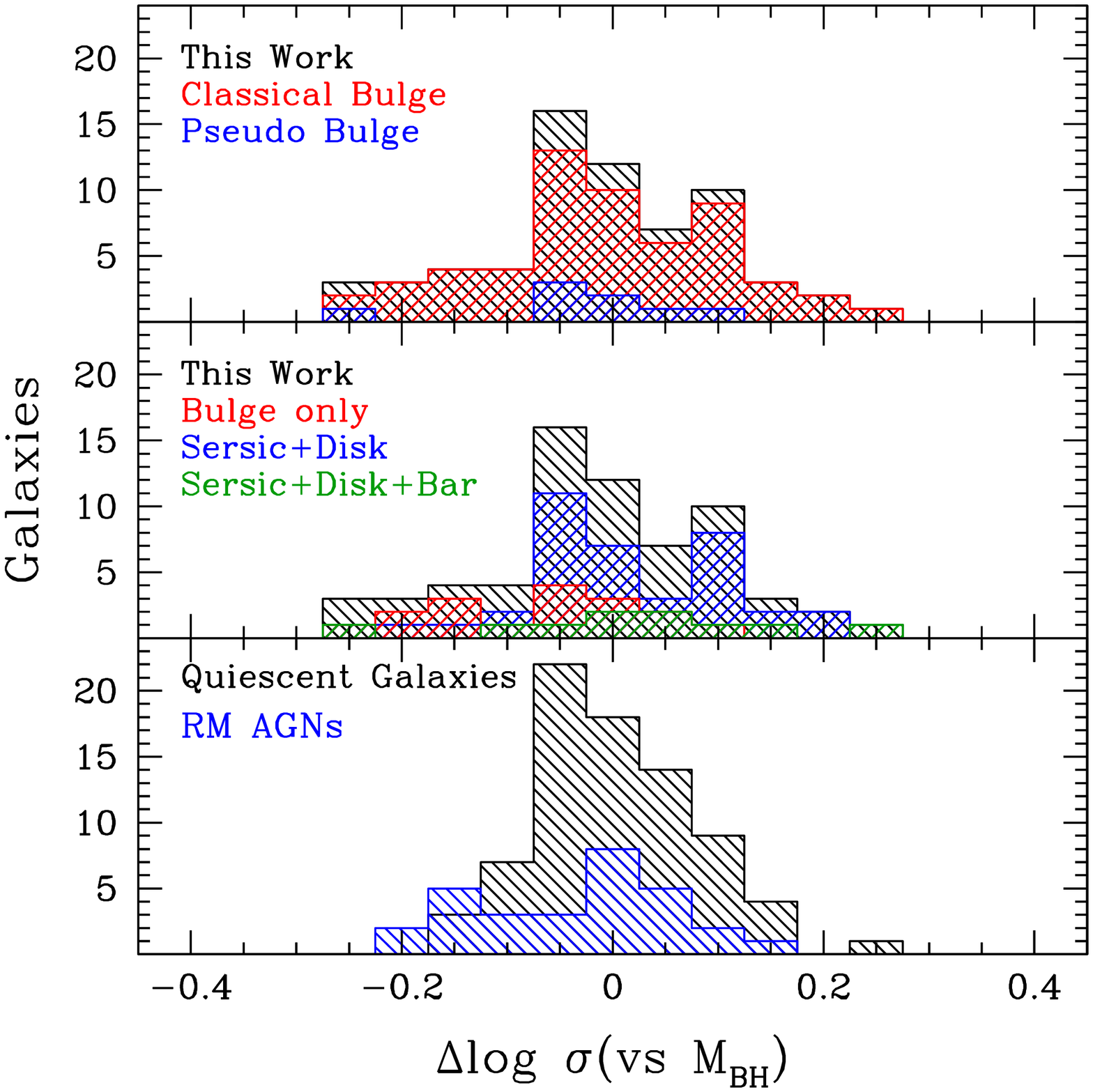}
\caption{
\mbh-\s~relation.
{\bf Left:}
\mbh-\s~relation for our sample (red open pentagons),
reverberation-mapped AGNs \citep[blue;][]{woo15}, and a sample of quiescent local galaxies 
\citep[black;][with the black dashed line being their best fit]{mcc13}.
The error on the BH mass for our sample is 0.4 dex
and shown as a separate point with error bar in the legend, to reduce confusion of data points.
For our sample, the stellar velocity dispersion was determined
within the spheroid effective radius according to Equation 1.
We assume a nominal
uncertainty of the stellar velocity dispersion of 0.04 dex.
The reverberation-mapped AGNs have $\sigma$ values derived 
from single apertures; $\sigma$ for the quiescent galaxies is 
determined similar to Equation 1, but within the galaxy effective radius
(see Section~\ref{sub:svd} for further comparison and discussion).
{\bf Right:}
Distribution of residuals with respect to the fiducial local
\mbh-\s~relation \citep[][Table~\ref{fits_relations}]{mcc13}. The lower panel shows
literature data (reverberation-mapped AGNs from \citet{woo15} in blue;
quiescent galaxies from \citet{mcc13} in black),
The middle panel shows our sample, the full sample in black, 
with different colors corresponding to different host-galaxy morphologies as 
indicated. The upper panel is the same as the middle panel,
but distinguishing between classical and pseudo bulges.}
\label{mbhsigma}
\end{figure*}

In Fig.~\ref{mbhsigma}, we show the resulting \mbh-$\sigma$ relation,
as well as the offset from the fiducial relation.
Overall, our sample follows the same \mbh-$\sigma$ relation as that
of reverberation-mapped AGNs as well as that of quiescent galaxies.
Our sample covers a small dynamical range in BH mass (6.7$<$$\log$\mbh$<$8.2), 
mainly due to the fact that we selected low-luminous Seyfert galaxies
with lower mass BHs to enable $\sigma$ measurements.
Considering the uncertainties of \mbh~of 0.4 dex,
we cannot independently determine the slope of the local relationship.
The sample size is, however, sufficient to determine
the zero point and scatter of the distribution around the local
relationship, assuming a choice of the slope.
Thus, when fitting a linear relation to the data of the form

\begin{eqnarray}
\log (M_{\rm BH}/M_{\odot}) = \alpha + \beta
\log(\sigma/200\,\rm km\,s^{-1})
\end{eqnarray}

we keep the value of $\beta$ fixed to the corresponding relationships of
quiescent galaxies (5.64 for \cite{mcc13} and 4.38 for \citet{kor13})
or reverberation mapped AGNs \citep[][3.97]{woo15}.
The results are summarized in Table~\ref{fits_relations}.
Both zero point and intrinsic scatter are comparable
to that of the quiescent galaxies, within the uncertainties,
consistent with the hypothesis that the \mbh~estimates we adopt do not
introduce significant uncertainty in addition to the estimated one.
Furthermore, biases based on different selection functions
(reverberation mapped AGNs selected based on variability;
quiescent galaxies selected based on the ability to resolve the BH
gravitational sphere of influence; our sample selected based on 
H$\beta$ line width) can be considered negligible.

When probing dependencies on host-galaxy morphology, 
we find that barred galaxies (comprising 17\% of the sample)
do not lie preferentially off the 
\mbh-\s~relation, in agreement with studies by \citet[e.g.,][]{gra08,ben09,bei12}
(see, however, \citealt{gra09}).
Also, neither merging galaxies (6\%) nor pseudo bulges (12\%)
form particular outliers from the relation.
This statement also holds true
when considering all candidate pseudo bulges 
(i.e.,~objects with spheroid with S{\'e}rsic index $n<2$; a total of 43\% of the sample).
In fact, candidate pseudo bulges show an even smaller scatter in the relation.

While the latter is in agreement with some studies \citep[e.g.,][]{kor01,gu09},
more recent studies suggest the opposite \citep{hu08,gre10,korb11,kor13,ho14}.
Pseudo bulges, characterized by nearly exponential light profiles, ongoing star
formation or starbursts, and nuclear bars, are believed to have evolved secularly
through dissipative processes rather than mergers \citep{cou96,kor04},
unlike their classical counterparts.

Given the sample of Seyfert-1 galaxies comprised of a majority
of late-type galaxies and the small fraction of mergers, our results
are consistent with secular evolution, driven by disk or bar instabilities
and/or minor mergers, growing both
BHs through accretion and spheroids
through a re-distribution of mass from disk to bulge
\citep[e.g.,][]{cro06,par09,jah09,cis11,ben10,ben11b,sch13}.
We can only speculate here that the smaller scatter exhibited by pseudo bulges
might in fact be explained by a synchronizing
effect that secular evolution has on the growth of BHs and bulges,
growing both simultaneously at a small but steady rate.
Major mergers, on the contrary, believed to create classical bulges
(and elliptical galaxies) are a more stochastic and dramatic phenomenon
with episodes of strong BH and bulge growth
that can be out of sync due to the different time scales involved
for growing BH and bulge in a major merger
\citep[e.g.,][]{hop12}.
We cannot directly probe the latter with our data --
as already mentioned, galaxies with obvious signs of interactions and mergers
do not form particular outliers from the relation, but this is based on a very small sample statistics of 6\% mergers.

However, we conclude with a note of caution: For one, when splitting our sample
into sub-samples (such as classical versus pseudo bulges), the results
suffer from small sample statistics.  Second, our morphological
classification relies on ground-based SDSS images with inherent
limitations such as limited depth and resolution.  For example, the
kpc-scale of bars and bulges is comparable to the spatial resolution
of the SDSS images (a $\sim1.5\arcsec$ PSF corresponds to 1.5 kpc at
z=0.05), significant given the presence of the bright point source in
AGNs.  The seeing in the ground-based SDSS images might also cause the
host galaxies to appear rounder than they actually are.  Likewise,
faint tidal features indicative of merger events might have been
missed in these shallow images.  Deeper and higher spatial-resolution
images are required to further test the connection between bars and
bulges and BHs.

\begin{deluxetable}{llccc}
\tabletypesize{\scriptsize}
\tablecolumns{2}
\tablewidth{0pc}
\tablecaption{Comparison between different definitions for the stellar velocity dispersion}
\tablehead{
\colhead{Ratio} & \colhead{Mean}\\
\colhead{(1)} & \colhead{(2)}}
\startdata
$\sigma_{\rm spat, 0.5 reff}$/$\sigma_{\rm spat, reff}$ & 0.98$\pm$0.01\\
$\sigma_{\rm spat, 0.25 reff}$/$\sigma_{\rm spat, reff}$  & 0.99$\pm$0.02\\
$\sigma_{\rm spat, reff, sini}$/$\sigma_{\rm spat, reff}$$^a$  & 1.31$\pm$0.09\\
$\sigma_{\rm spat, 0.5 reff, sini}$/$\sigma_{\rm spat, reff}$ & 1.14$\pm$0.06\\
$\sigma_{\rm spat, 0.25 reff, sini}$/$\sigma_{\rm spat, reff}$  & 1.05$\pm$0.04\\
$\sigma_{\rm spat, reff,galaxy}$/$\sigma_{\rm spat, reff}$$^a$  & 0.99$\pm$0.01\\
$\sigma_{\rm spat, 0.5 reff,galaxy}$/$\sigma_{\rm spat, reff}$  & 0.96$\pm$0.01\\
$\sigma_{\rm spat, 0.25 reff,galaxy}$/$\sigma_{\rm spat, reff}$ & 0.98$\pm$0.02\\
$\sigma_{\rm spat, reff,galaxy, sini}$/$\sigma_{\rm spat, reff}$  & 1.43$\pm$0.09\\
$\sigma_{\rm spat, 0.5 reff,galaxy, sini}$/$\sigma_{\rm spat, reff}$  & 1.18$\pm$0.06\\
$\sigma_{\rm spat, 0.25 reff,galaxy, sini}$/$\sigma_{\rm spat, reff}$  & 1.07$\pm$0.04\\
$\sigma_{\rm spat, SDSS}$/$\sigma_{\rm spat, reff}$$^a$  & 0.97$\pm$0.01\\
$\sigma_{\rm ap, reff}$/$\sigma_{\rm spat,reff}$ & 1.01$\pm$0.02
\enddata
\tablecomments{
Comparison between the different definitions for the stellar velocity dispersion
used in the literature.
Col. (1): Ratio between a given definition of $\sigma$ 
(see text for details) and the fiducial $\sigma_{\rm spat, reff}$ used
throughout the paper (Eq. 1).
Col. (2): Mean and uncertainty of the ratio.\\
$^a$ Shown in Fig.~\ref{fig_sigmacompare}.\\
}
\label{table_sigmacompare}
\end{deluxetable}

\begin{deluxetable*}{llcccc}
\tabletypesize{\scriptsize}
\tablecolumns{7}
\tablewidth{0pc}
\tablecaption{Fits to the Local \mbh-\s~Relations}
\tablehead{
\colhead{Sample} & \colhead{$\alpha$} & \colhead{$\beta$} & \colhead{Scatter} & \colhead{Offset} & \colhead{Reference}\\
\colhead{(1)} & \colhead{(2)} & \colhead{(3)}  & \colhead{(4)} & \colhead{(5)} & \colhead{(6)}}
\startdata
Quiescent Galaxies (72) & 8.32$\pm$0.05 & 5.64$\pm$0.32 & 0.38 & & \citealt{mcc13}$^a$\\
Quiescent Galaxies (51)  & 8.49$\pm$0.05 & 4.38$\pm$0.29 & 0.29 & & \citealt{kor13}\\
Reverberation-mapped AGNs (29)  & 8.16$\pm$0.18 & 3.97$\pm$0.56 & 0.41$\pm$0.05 & & \citealt{woo15}\\
AGNs (66)              & 8.38$\pm$0.08 & 5.64 (fixed) & 0.43$\pm$0.09 & -0.01$\pm$0.01 & this paper\\
AGNs (66)              & 8.20$\pm$0.06 & 4.38 (fixed) & 0.25$\pm$0.10 & 0.06$\pm$0.01 & this paper\\
AGNs (66)              & 8.14$\pm$0.06 & 3.97 (fixed) & 0.19$\pm$0.10 & 0.02$\pm$0.01 & this paper
\enddata
\tablecomments{
Fits to the \mbh-\s~relation, $\log (M_{\rm BH}/M_{\odot}) = \alpha +
\beta \log (\sigma / 200 {\rm km\,s}^{-1})$  
Col. (1): Sample and sample size in parenthesis.
Col. (2): Mean and uncertainty on the best fit intercept.
Col. (3): Mean and uncertainty on the best fit slope.
Col. (4): Mean and uncertainty on the best fit intrinsic scatter.
Col. (4): Mean and uncertainty of offset from fiducial relation of either \citet{mcc13} (slope fixed to 5.64), \citet{kor13}
(slope fixed to 4.38), or \citet{woo15} (slope fixed to 3.97).
Col. (5): References for fit. 
Note that the quoted literature uses FITEXY with a uniform prior on the intrinsic scatter,
so our fits assume the same.\\
$^a$ Relation plotted as dashed lines in Fig.~\ref{mbhsigma} and
used as fiducial relation when calculating residuals.}
\label{fits_relations}
\end{deluxetable*}

\newpage
\section{SUMMARY}
\label{sec:summary}
To understand the origin of the scaling relations between the mass of
the central supermassive BH and the properties of the host galaxy,
studies relying on type-1 active galaxies probe the evolution of these
relations.  The robust determination of slope, scatter, and
dependencies of a baseline consisting of a comparable sample of type-1
AGNs in the local Universe is essential to minimize biases before any
conclusions about their evolution can be drawn.  We here create a
local baseline of the \mbh -- stellar-velocity
dispersion ($\sigma$) relation for a homogeneously selected sample of
66 Seyfert-1 galaxies in the local Universe (0.02 $< z <$ 0.09)
selected from SDSS-DR6 based on BH mass (\mbh$>10^7 M_{\odot}$),
Combining high S/N long-slit Keck spectra with SDSS images yields
\mbh~using the virial method and stellar-velocity dispersion
($\sigma_{\rm spat, reff}$) from spatially-resolved kinematics.
Our results can be summarized as follows.

\begin{enumerate}

\item The majority of host galaxies (77\%) are classified as Sa or later
with roughly one-quarter of those showing evidence for a bar.  This
high fraction is also reflected in prominent rotation curves with a
maximum velocity of 100-200 km\,s$^{-1}$ in the majority of kinematic
measurements (60\%).  The majority of spiral galaxies (28 objects;
55\%) have spheroid components with S{\'e}rsic index $n<2$. Of these
candidate pseudo bulges, eight can be considered as definite, when applying
the classification criteria given by \citet{kor13} that are measurable
with our data. The minority (6\%) shows signs of merging or
interacting galaxies, comparable to quiescent galaxies in the local
Universe.  These host-galaxy morphologies are typical for a sample of
Seyfert-1 galaxies and suggest BH accretion being dominated by secular
processes.

\item {We use our spatially resolved kinematic data to reproduce various 
measurements of the stellar-velocity second moment, as used in the
literature.  The derived quantities differ significantly across
definitions, in particular when considering effects of inclination,
inclusion of a kinematically cold, rotating disk, and an AGN power law
continuum swamping the stellar absorption lines in the center. The average
differences can be as much as 40\%, highlighting the importance of
using self-consistent definitions when comparing samples.
We consider the definition of $\sigma$ used in this paper
as derived from spatially resolved measurements and integrated out to the bulge
effective radius the most robust value for studies of the
\mbh-\s~relation, excluding disk rotation and AGN contamination.}

\item {Our Seyfert-1 galaxy sample follows the same
\mbh-\s~relation as the one of reverberation-mapped
AGNs as well as that of quiescent galaxies,
with the same intercept and scatter. This is consistent with
the hypotheses that the uncertainty of the single epoch \mbh~estimates
we adopt is not significantly underestimated, and that the
broad-emission line selection does not introduce a significant bias
with respect to the criteria used in comparison samples
(e.g., variability or BH sphere of influence).}

\item {Neither barred galaxies, merging galaxies, nor galaxies with
pseudo bulges seem to be significant outliers of the relation, at
variance with recent predictions in the literature \citep{kor13}.
However, this conclusion is based on small sample statistics and relies
on low-resolution ground-based SDSS images which make a morphological
classification challenging in the presence of a bright AGN point
source. Larger samples with higher quality data are needed to further
test any correlations between detailed host-galaxy morphology and 
BH mass.}

\end{enumerate}

\medskip

In the next paper of this series we will discuss the other scaling
relations, namely with spheroid luminosity and stellar mass as well as
host-galaxy luminosity and stellar mass.

\acknowledgments
We thank the anonymous referee for valuable comments helping to improve the paper.
VNB thanks Aaron Barth, St{\'e}phane Courteau, Eric Emsellem, and Stefanie Komossa for discussions.
VNB acknowledges assistance from a
National Science Foundation (NSF) Research at Undergraduate
Institutions (RUI) grant AST-1312296.
Note that findings and conclusions do not necessarily represent views of the
NSF. VNB and TT acknowledge 
support for program number  HST-AR-12625.11-A, provided by NASA through
a grant from the Space Telescope Science Institute, which is operated by the
Association of Universities for Research in Astronomy, Incorporated, under
NASA contract NAS5-26555.
TT acknowledges support from the Packard Foundations in the form of a Packard Fellowship.
DP acknowledges support through the EACOA Fellowship from The East Asian Core Observatories Association, which consists of the
National Astronomical Observatories, Chinese Academy of Science (NAOC), the National Astronomical Observatory of Japan (NAOJ), Korean
Astronomy and Space Science Institute (KASI), and Academia Sinica Institute of Astronomy and Astrophysics (ASIAA).
JHW acknowledges support by the National Research Foundation of Korea
(NRF) grant funded by the Korea government (2012-006087 and
2010-0027919).
Data presented in this paper were obtained at the W. M. Keck Observatory, 
which is operated as a scientific partnership among Caltech, the University of California, and NASA. 
The Observatory was made possible by the generous financial support of the W. M. Keck Foundation. 
The authors recognize and acknowledge the very significant cultural role and reverence that the summit of 
Mauna Kea has always had within the indigenous Hawaiian community. 
We are most fortunate to have the opportunity to conduct observations from this mountain. 
This research has made use of the public archive of the Sloan Digital Sky Survey and the NASA/IPAC Extragalactic Database (NED) 
which is operated by the Jet Propulsion Laboratory, California Institute of Technology, 
under contract with the National Aeronautics and Space Administration.

{\it Facilities:} \facility{Keck:I (LRIS)}

\appendix
\section{BROAD H$\beta$ EMISSION-LINE FITTING}
\label{appendix1}
\begin{figure*}[ht!]
\includegraphics[scale=0.45]{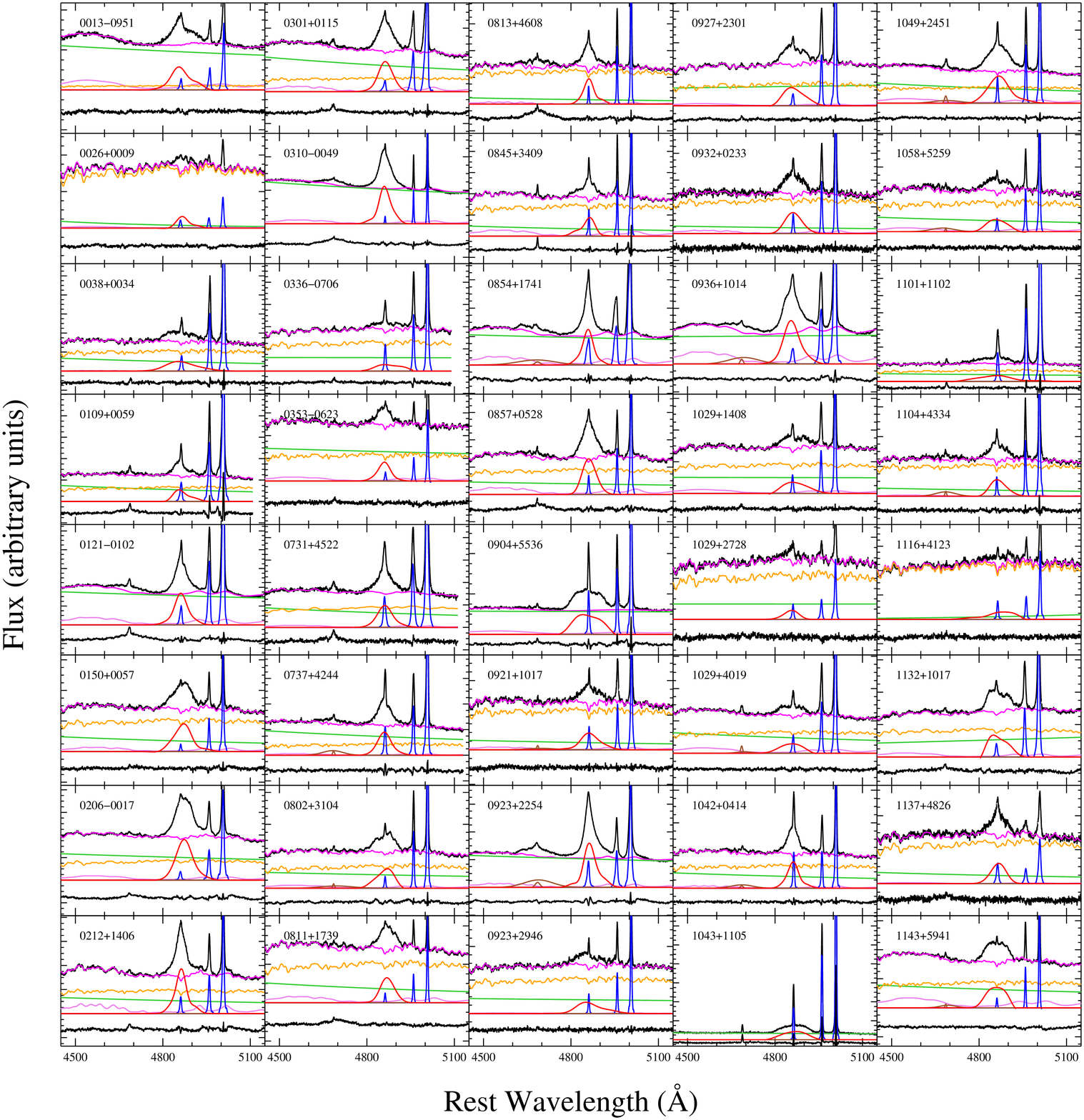}
\caption{
Determination of the second moment of the broad H$\beta$ emission
using a multi-component spectral decomposition.
In the upper region, the observed spectrum is shown in black
with the best-fit model consisting of the continuum, FeII, and host galaxy starlight in magenta.
Below, the best-fit to the power-law continuum is
shown in green with the stellar spectrum in yellow.
Further below are the narrow lines of H$\beta \lambda$4861, and [OIII]
$\lambda\lambda$4959,5007 in blue, the broad and narrow components of HeII $\lambda$4686 in brown
(only included if blended with the broad H$\beta$), 
the broad component of H$\beta$ in red and the FeII contribution in purple. 
Finally, the residuals are shown in black (arbitrarily shifted downward for clarity).}
\label{hbeta1}
\end{figure*}

\begin{figure*}[ht!]
\includegraphics[scale=0.4]{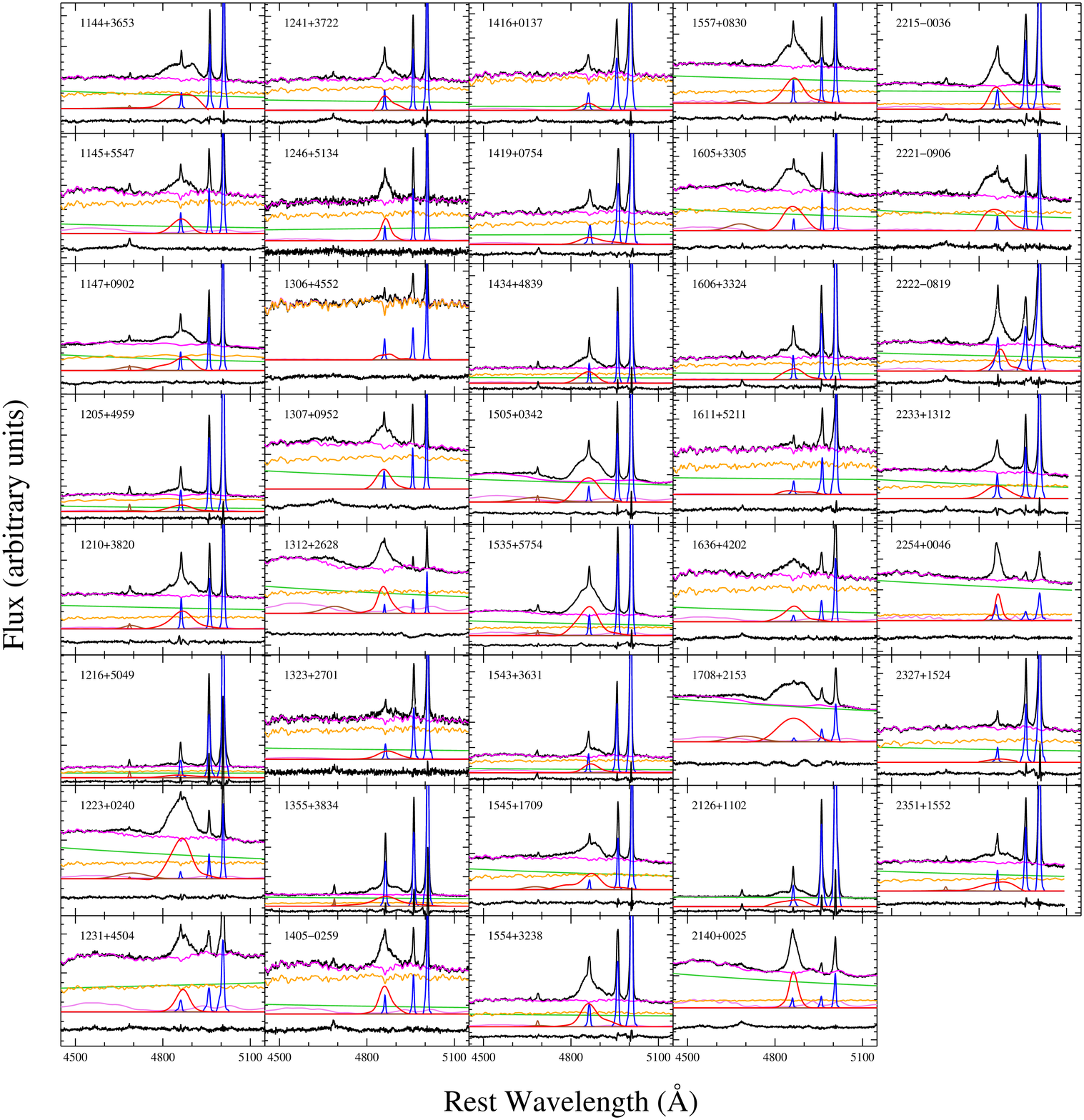}
\caption{Fig.~\ref{hbeta1} continued}
\label{hbeta2}
\end{figure*}

Figs.~\ref{hbeta1} and~\ref{hbeta2} show fits to the broad H$\beta$ emission line for 79 objects
for which we were able to measure \mbh.

\newpage
\section{NOTES ON INDIVIDUAL OBJECTS}
\label{appendix2}
One object in our sample, 1223+0240 also known as MRK50,
is also in the reverberation-mapped AGN sample and its BH mass has been estimated
both in a traditional way \citep{bar11},
$\log M_{\rm BH}/M_{\odot}=7.51\pm0.05$
(and that is also the value used when plotting the reverberation-mapped AGNs;
note that the quoted uncertainty does not include the uncertainty on the virial factor)
as well as using dynamical modeling \citep{pan12},
$\log M_{\rm BH}/M_{\odot}=7.57^{+0.44}_{-0.27}$.
Both values are within the uncertainties of the BH mass determined here via the virial method of 
$\log M_{\rm BH}/M_{\odot}=7.25\pm0.4$.
\citet{pan12} infer that the geometry of the BLR is a nearly face-on thick disk,
with a potential net inflow or outflow.
MRK50 is bulge dominated \citep{mck90,kos11} 
and \citet{ho14} classify its bulge as classical bulge which matches our classification.
Note that Fig.~\ref{mbhsigma} includes MRK50 twice, once amongst the reverberation-mapped AGNs
with the reverberation-mapped MBH, once amongst our local sample.

1535+5754, aka MRK290, was included in the Lick AGN Monitoring Program (LAMP 2008),
but did not exhibit strong variations, so no BH mass was derived.
\citet{ben13} classify the host galaxy as
an early-type spiral galaxy (Sa-Sab) at a relatively low inclination,
based on HST images. However, this classification is questionable.
\citet{deo06} describe the host galaxy as ``probably elliptical'', but mistakenly classified as unbarred spiral
previously \citep{cren03}. We classified MRK290 as elliptical galaxy with a classical bulge.

1216+5049 is known as MRK1469 and was classified as a highly inclined spiral galaxy
with only large-scale dust lanes visible in \citet{deo06}, in agreement with our classification.

\end{document}